\documentclass[aps,prb,twocolumn,floatfix,showpacs]{revtex4}
\usepackage{graphicx}

\begin{document}

\title{Is manganese-doped diamond a ferromagnetic semiconductor?}
\author{Steven C. Erwin and C. Stephen Hellberg}
\affiliation{Center for Computational Materials Science, Naval Research Laboratory,
Washington, D.C. 20375}
\date{\today}

\begin{abstract}
We use density-functional theoretical methods to examine the recent
prediction, based on a mean-field solution of the Zener model, that
diamond doped by Mn (with spin $S$=5/2) would be a dilute magnetic
semiconductor that remains ferromagnetic well above room temperature.
Our findings suggest this to be unlikely, for four reasons: (1)
substitutional Mn in diamond has a low-spin $S$=1/2 ground state; (2)
the substitutional site is energetically unfavorable relative to the
much larger ``divacancy'' site; 3) Mn in the divacancy site is an
acceptor, but with only hyperdeep levels, and hence the holes are
likely to remain localized; (4) the calculated Heisenberg couplings
between Mn in nearby divacancy sites are two orders of magnitude
smaller than for substitutional Mn in germanium.
\end{abstract}

\maketitle

\section{Introduction}

A large class of dilute magnetic semiconductors (DMS) is based on
manganese doping of III-V or group-IV hosts having the zincblende,
wurzite, or diamond structure.  Formally, Mn is either a single or
double acceptor, depending on whether it substitutes for a group-III
or -IV atom,
respectively.\cite{ohno_appl_phys_lett_1996a,park_science_2002a} Aside
from this difference, there are no simple guidelines for predicting
how the magnetic behavior of the resulting DMS depends on the choice
of host semiconductor.  Elucidating such guidelines would be of great
practical interest for efforts to control and optimize the magnetic
properties of DMS materials. In particular, it would be very helpful
to understand how the resulting Curie temperatures depend on the
choice of host semiconductor.

An important and early contribution was made by \mbox{Dietl}, who used
a mean-field solution of the Zener model to predict the Curie
temperatures for Mn doping of a wide range of II-VI, III-V, and
group-IV host
semiconductors.\cite{dietl_science_2000a,dietl_phys_rev_b_2001a} Mn
ions with localized spin $S$=5/2 were assumed to substitute on the
cation site, and to interact by an indirect exchange interaction
mediated by holes.  Curie temperatures were calculated for fixed Mn
content $x=5\%$ and hole concentration $p=3.5\times10^{20}$ cm$^{-3}$
using the mean-field result
\begin{equation}
k_BT_C=x_\mathrm{eff}N_0S(S+1)\beta^2A_Fm^\ast k_F/12h^2.
\end{equation}
In this equation, the most direct dependence of $T_C$ on the host
semiconductor comes from the density per unit volume of
cation sites $N_0$, which varies with the host lattice constant as
$1/a^3$.  A much weaker dependence on the host arises from the
density-of-states effective mass $m^\ast$ and the Fermi wavevector
$k_F$.  The remaining variables, i.e.~the $p$-$d$ exchange integral $\beta$ and the Fermi-liquid parameter
$A_F$, were assumed not to depend on the host.

The full numerical evaluation of Eq.~(1) for different host semiconductors\cite{dietl_phys_rev_b_2001a}
leads to values of $T_C$ that indeed track the prefactor $1/a^3$ quite
closely.\cite{erwin_unpub_2003a}  For example, the hosts with
the largest lattice constants (CdTe, InAs, Ge) 
lead to the smallest predicted Curie temperatures (below $\sim$75 K).
Likewise, the hosts with the smallest lattice constants (ZnO, GaN, C) lead
to the largest 
predicted values of $T_C$ (well above room temperature).

Diamond has the smallest lattice constant of all the semiconductors
($a=3.56$ \AA) and, accordingly, the highest predicted Curie
temperature ($T_C\approx$ 470 K).  These extreme values make diamond a
logical choice for examining some of the assumptions underlying the
predictions of high Curie temperatures for hosts having small lattice
constants. In this paper, we use first-principles theoretical methods
to test the assumption that Mn occupies the substitutional site, with
spin $S=5/2$, in the diamond lattice.  We find that substitutional Mn
has a low-spin $S$=1/2 ground state in diamond. Moreover, because of
diamond's small lattice constant, the substitutional site is
energetically unfavorable relative to the much larger ``divacancy
site,'' denoted $V_2$, which we identify as the more likely Mn
impurity location. We analyze the electronic and magnetic structure of
this divacancy Mn impurity in diamond, and show that it is unlikely to
lead to magnetic ordering at any reasonable temperature.

\begin{figure}[b]
\resizebox{7cm}{!}{\includegraphics{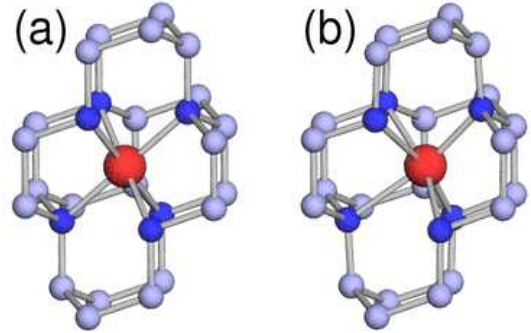}}
\caption{(a) Ideal and (b) completely relaxed geometry of the Mn impurity in a
diamond divacancy site.}
\end{figure}

\section{Methods}

Most of the results reported in this article are for isolated Mn
impurities, which we simulate using a supercell of 64 carbon atoms in
the diamond structure.  All atomic positions were relaxed using total
energies and forces calculated within the generalized-gradient
approximation to density-functional theory (DFT), as implemented in
{\sc vasp}.\cite{kresse_phys_rev_b_1993a,kresse_phys_rev_b_1996a}
Carbon and Mn ultrasoft pseudopotentials were used with a cutoff
energy of 286 eV.  For total energies, we used 2$\times$2$\times$2
Monkhorst-Pack sampling of the Brillouin zone, and
4$\times$4$\times$4 sampling for convergence checks. In the discussion of
the electronic structure of the isolated Mn impurity at a divacancy
site, eigenvalues at the zone center of a 128-atom supercell are
reported in order that degeneracies are properly represented.

Within the supercell formalism, the formation energy of a Mn impurity
is given by
\begin{equation}
E_{\rm form}[{\rm Mn}^q] = E_t[{\rm Mn}^q] 
                  - n_{\rm C} \mu_{\rm C}
                  - \mu_{\rm Mn}
                  + q E_F,
\end{equation}
where $E_t[{\rm Mn}^q]$ is the total energy of a supercell containing
$n_{\rm C}$ carbon atoms and one Mn impurity, with chemical potentials
$\mu_{\rm C}$ and $\mu_{\rm Mn}$, $q$ is the charge state of the Mn
impurity, and $E_F$ is the Fermi level. Since the host crystal is elemental,
the chemical potentials are simply the energy per atom in the
diamond phase of carbon and the ground-state $\alpha$Mn.

The electrical activity of a Mn impurity is determined by its
formation energy as a function of charge state.  Electrically active
defects will have more than one stable charge state within the host
band gap; the value of $E_F$ for which two such charge states have
equal formation energies is referred to as a ``charge transition
level.''  The charge transition level between the neutral and $q=+1$
states is the donor ionization energy, while the transition level
between neutral and $q=-1$ states is the acceptor ionization energy.
In practice, the total energy of a charged supercell must be
calculated by adding a uniform compensating background charge to the
supercell.  The long-range nature of the Coulomb interaction then
gives rise to a spurious Madelung-like contribution to the total
energy, which must be subtracted from the calculated total energy.
This was done using a standard approach in which a multipole expansion
of the defect charge (up to quadrupole order) was used to estimate the
interaction energy analytically and then subtract it
off.\cite{makov_phys_rev_b_1995a}

The magnetic interaction between Mn impurities in nearby divacancy
sites was calculated using larger supercells of 128 carbon atoms. No
relaxations were performed for these calculations; tests for a few
configurations confirmed that this approximation did not change the
results significantly. The interactions are represented by
numerically mapping the DFT total energies into the Heisenberg
form. In practice this amounts to computing the
difference in total energy between the parallel and antiparallel spin
alignment of two Mn impurities in a supercell.

\section{Energetics}

\begin{figure}
\resizebox{7cm}{!}{\includegraphics{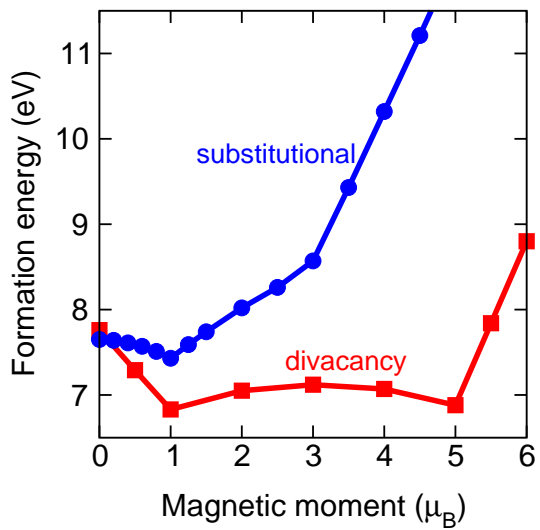}}
\caption{Formation energy of substitutional and divacancy Mn in
diamond, as a function of constrained magnetic moment.}
\end{figure}

The Mn impurity enjoys a special status as a dopant in dilute magnetic
semiconductors such as Mn$_x$Ga$_{1-x}$As and Mn$_x$Ge$_{1-x}$.  It
preferentially occupies the substitutional (cation) site, where it
serves two roles: it is electrically active as an acceptor, and it
contributes a localized spin $S=5/2$.  At sufficient Mn concentrations,
holes are created in the host valence band; these delocalized carriers
mediate the ferromagnetic interaction between the localized
spins.\cite{dietl_science_2000a,dietl_phys_rev_b_2001a}

GaAs and Ge have similar lattice constants, and since strain
contributes to the impurity formation energy, it is instructive to
note that the local strain around a substitutional Mn impurity, denoted
Mn$_{\rm sub}$, is very small for both hosts---the nearest neighbors
distort by less than 2\% from their ideal positions.  Tetrahedrally
coordinated interstitial Mn impurities, denoted Mn$_{\rm tet}$, have higher formation energies
(by 1--2 eV for neutral impurities); interestingly, the atomic volume
available at this site is the same as for the substitutional site, and
consequently very little distortion is created by the Mn interstitial
in either host.\cite{erwin_phys_rev_lett_2002a}

The
lattice constant of diamond is 37\% smaller than that of Ge, and so it
is not surprising that a substitutional Mn impurity creates a much
larger local strain in the lattice---within DFT the nearest-neighbor distortion
is 12\%. Because of the extreme stiffness of diamond, such a large
strain is expected to lead to a relatively high formation energy.
A similar distortion occurs for Mn in the tetrahedral interstitial site.

\begin{figure*}
\resizebox{15cm}{!}{\includegraphics{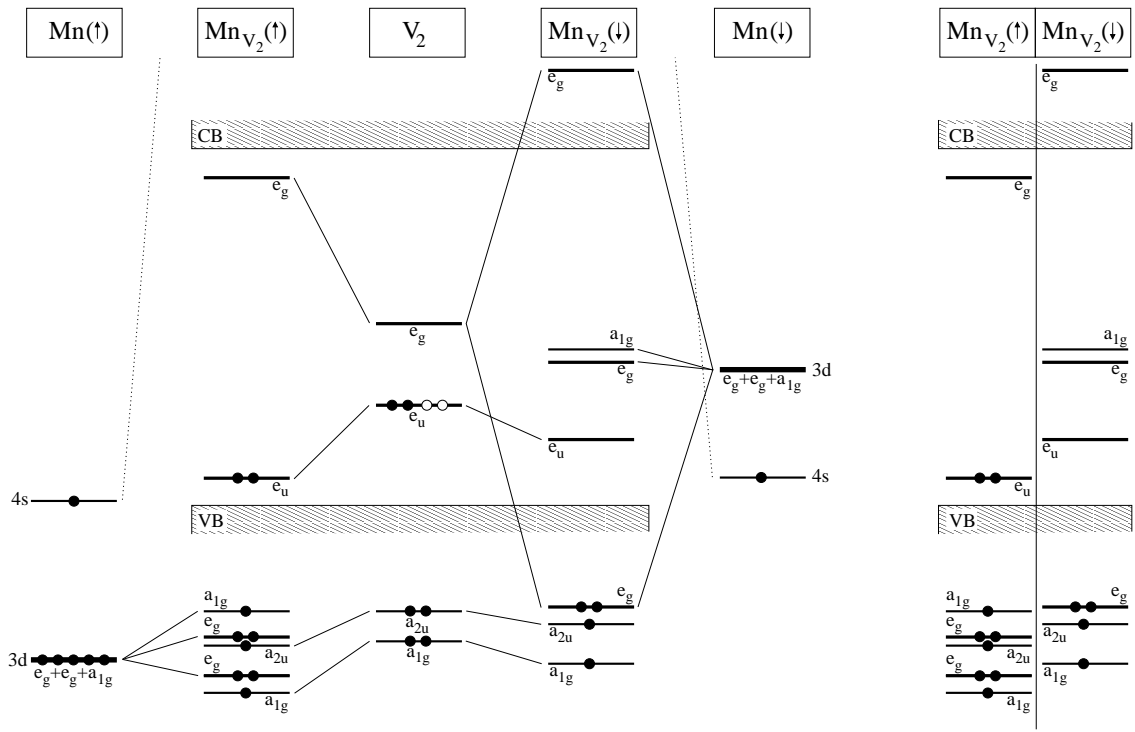}}
\caption{Energy-level diagram for divacancy Mn impurity in
diamond, Mn$_{V_2}$. Left panels show how majority $(\uparrow)$
and minority $(\downarrow)$ levels arise from hybridization of Mn
atomic states with diamond divacancy ($V_2$) states.  Right panel summarizes the
resulting energy levels.}
\end{figure*}

A simple way to relieve this strain is to place the Mn impurity at a
divacancy site---i.e.~to remove two neighboring carbon atoms and place
the Mn at their midpoint, as shown in Fig.~1(a). This doubles the
atomic volume available, leading to a much reduced local strain---only
2\% distortion of the nearest neighbors. Even in its ground-state
configuration, which includes a slight off-center relaxation of the Mn
ion as shown in Fig.~1(b), the local lattice distortion around this
Mn$_{V_2}$ impurity is very small.  It is already widely believed that
other large impurities---including Si, P, Ni, and Co---preferentially
occupy the divacancy site in
diamond;\cite{goss_phys_rev_lett_1996a,jones_appl_phys_lett_1996a,nadolinny_applied_magnetic_resonance_1997a,twitchen_phys_rev_b_2000a,iakoubovskii_phys_rev_b_2000a}
hence our prediction that Mn also prefers this site appears plausible.

To make the comparison between different sites quantitative we turn now to the formation
energies calculated within DFT. For neutral Mn impurities, we find the
divacancy site is strongly preferred (by 0.6 eV) to the substitutional
site; the interstitial site is much less favorable than either (by $\sim$10
eV). Figure 2 shows the formation energies for divacancy and substitutional Mn
as a function of magnetic moment, calculated using the
fixed-spin-moment method within DFT.  Even in the preferred divacancy site, the absolute
formation is almost an order of magnitude larger for substitutional Mn
in GaAs or Ge, suggesting that standard methods for attaining high
concentrations of magnetic dopants will probably not work for
diamond. 

The equilibrium magnetic moments $M$ in the substitutional and
divacancy sites are also quite different compared to substitutional Mn
in GaAs and Ge ($M=4$ and 3 $\mu_B$, respectively).  In the
substitutional site, Mn adopts a low-spin $M=1$ state, with an
extremely small energy gain (0.1 eV) relative to the spin-unpolarized
state. Hence, we anticipate that even if Mn could be forced into the
substitutional site, its magnetic moment may not be sufficiently
stable to facilitate magnetic ordering.  In the divacancy site two nearly
degenerate moments are found ($M=1$ and 5 $\mu_B$) with an energy
difference of only 0.05 eV; the formation energy vs.~magnetic moment
is symmetrical about $M=$ 3 $\mu_B$.  This peculiar feature of the Mn$_{V_2}$
impurity will be explained in Section IV.

\section{Electronic structure}

The Mn$_{V_2}$ impurity has an energy-level structure that arises from
the relatively weak interaction of atomic Mn levels and host divacancy
levels, as shown in Fig.~3.  The host divacancy defect itself has been
extensively studied both theoretically and experimentally, and is well
understood.\cite{watkins_phys_rev_1965a,humphreys_j_phys_c_1983a,pesola_phys_rev_b_1998a,coomer_physica_b_1999a}
The removal of two carbon atoms from the host lattice leads to a
defect with $D_{3d}$ symmetry, and thus the six available
dangling-bond orbitals must form linear combinations that transform
under the irreducible representations $a_{1g}$, $a_{2u}$, $e_u$, and
$e_g$.\cite{nomenclature} For the neutral divacancy, six electrons are available to fill
these levels. In an early theoretical study, Coulson and
Larkins\cite{coulson_j_phys_chem_solids_1969a} predicted the ground
state one-electron configuration to be $a_{1g}^2a_{2u}^2e_u^2$.  Later
calculations confirmed this one-electron level ordering for both
silicon and diamond, and we find it here too, as shown.

In Fig.~3 we consider separately the interaction of the majority and
minority Mn levels with the four divacancy levels. Under the $D_{3d}$
crystal field, the 3$d$ level of Mn splits into one $a_{1g}$ and two
$e_g$ levels.  Hence, hybridization is allowed only between Mn and
$V_2$ levels both having either $a_{1g}$ or $e_g$ symmetry, but no
mixing is allowed between Mn levels and $V_2$ levels having either
$a_{2u}$ or $e_u$ symmetry.  Among the calculated majority
Mn$_{V_2}(\uparrow)$ levels we find very little mixing within any
level, and therefore each has a clear parentage from either Mn or
$V_2$ orbitals, as indicated in Fig.~3.  For the minority Mn$_{V_2}(\downarrow)$
levels there is strong mixing between the empty $V_2$ $e_g$ level and
the empty Mn $e_g(\downarrow)$ level, so that the occupied bonding
combination falls well below the diamond valence-band maximum (VBM).
Considering for the moment only those levels that sit below the VBM,
we find that seven electrons occupy majority levels and four electrons
occupy minority levels. Hence, these low-lying levels carry a net ``core''
magnetic moment $M_{\rm core}=3 \mu_B$.

\begin{figure}
\resizebox{7cm}{!}{\includegraphics{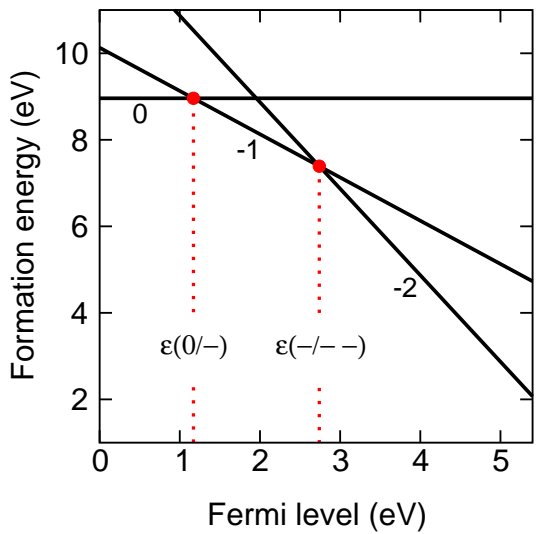}}
\caption{Formation energy of divacancy Mn impurity in diamond, 
as a function of the Fermi level.  Two charge-transition levels fall
within the diamond band gap, as indicated.}
\end{figure}

The highest occupied molecular orbital (HOMO) of the Mn$_{V_2}$
impurity has $e_u$ symmetry and, because it cannot mix with Mn states,
is purely derived from the $V_2$ dangling-bond orbitals.  The two
electrons in this fourfold level form a spin triplet ($M_{\rm HOMO} =
$ 1 $\mu_B$) as shown, but the system is energetically indifferent to
whether they are aligned parallel or antiparallel to the core spin
(Fig.~3 shows the parallel alignment; when the alignment is made
antiparallel the spin-up and spin-down HOMO levels simply switch
positions). 

Because of this weak coupling between the HOMO spin and the core spin,
the total magnetic moment of Mn$_{V_2}$ can assume two possible
low-energy values: either $M_{\rm core}+M_{\rm HOMO}=$ 5 $\mu_B$ or
$M_{\rm core}-M_{\rm HOMO}=$ 1 $\mu_B$ for parallel or antiparallel
alignment, respectively. Moreover, there is an energy penalty for violating
Hund's Rule by forcing the HOMO into a spin-singlet state; this is the
origin of the energy increase that occurs when the total moment is
constrained to 3 $\mu_B$ (the core moment alone). These two
features---the weak core-HOMO spin coupling, and the Hund's Rule
penalty for constraining the HOMO to a spin singlet---explain the
unusual appearance, shown earlier in Fig.~2, of the Mn$_{V_2}$
formation energy vs.~total magnetic moment.

\section{Electrical activity}

In Fig.~4 we show the calculated formation energies of Mn$_{V_2}$ for
charge states that are stable within the experimental 5.4-eV band gap
of diamond.  There are three such stable charge states ($q=0,-1,-2$),
making Mn$_{V_2}$ formally a double acceptor. In principle this
is similar to the case of substitutional Mn in Ge, which is also a
double acceptor, and might suggest similar electrical activity for
Mn$_{V_2}$.

\begin{figure}[b]
\resizebox{7cm}{!}{\includegraphics[clip]{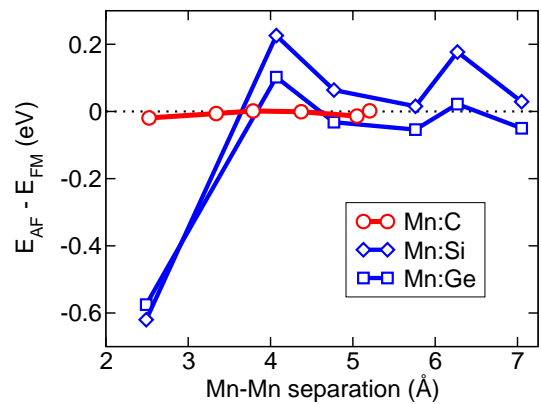}}
\caption{Heisenberg spin-coupling constant between two Mn impurities
in C (diamond), Si, and Ge host crystals.  For diamond, Mn atoms are
at divacancy sites; for Si and Ge, Mn atoms are on substitutional sites.}
\end{figure}

Quantitatively, however, the two cases are quite different. The first
two acceptor ionization energies for Mn in Ge have been measured
experimentally to be 160 and 370 meV, relative to the valence-band
edge.\cite{sze1981a} For Mn concentrations of several percent, the
width of the resulting impurity band is sufficient to create overlap
with the valence band and thus to allow the holes to delocalize.
Likewise, Mn in GaAs is a single acceptor with a measured acceptor
level at 113 meV,\cite{schneider_phys_rev_lett_1987a} which is
accurately given by DFT calculations as 100 meV.
\cite{erwin_phys_rev_lett_2002a}

On the other hand, the predicted first acceptor ionization energy for
Mn$_{V_2}$ in diamond is an order of magnitude larger,
$\epsilon(0/-)=$ 1.2 eV, suggesting that the holes will remain
localized on the impurity sites for any reasonable Mn
concentration. For this reason alone, it appears very unlikely that
there will be sufficient carriers to mediate a ferromagnetic
interaction between the localized spins of Mn$_{V_2}$ impurities.

\section{Magnetic interactions}

The weak coupling between core and HOMO spins in the isolated
Mn$_{V_2}$ impurity suggests that the effective coupling between spins
of two impurities will also be weak.  This expectation is confirmed in
Fig.~5, where we show the results of our DFT calculations for (twice)
the Heisenberg coupling constant between two Mn$_{V_2}$ impurities in
a 128-atom diamond supercell. For comparison, we show the results of
similar calculations for substitutional Mn in Ge and Si.

The results for substitutional Mn in Ge and Si show a similar trend.
When the two Mn are nearest neighbors, there is a strong preference
for antiparallel spin alignment. At larger separations the
interactions vary in sign (Ge) and/or magnitude (Si), a consequence of
the crystallographic anisotropy that arises from the strong $d$
character of the acceptor wavefunction in substitutional Mn. The
largest ferromagnetic coupling strengths are in the range 0.1$-$0.2 eV.
(For Si there is a shift of all the couplings toward more favorable
ferromagnetic interactions; since Mn-doped Ge is known to be
ferromagnetic this suggests that Mn-doped Si might also be---perhaps with
a higher Curie temperature).

The coupling between two Mn$_{V_2}$ impurities in diamond is
substantially weaker: the largest ferromagnetic couplings are of order
1 meV, two orders of magnitude smaller than for Mn in Ge or Si. Hence,
any ferromagnetically ordered phase involving Mn$_{V_2}$ impurities
would be characterized by a Curie temperature at most on this
scale, i.e.~of order 10 K.

\section{Summary}

We have shown theoretically that the Mn impurity in diamond is
energetically more favorable in the divacancy site than in the
substitutional site. The magnetic properties of the divacancy Mn
impurity show nearly degenerate low-spin and high-spin configurations;
the degeneracy arises from the weak coupling of separate core and
valence moments.  Divacancy Mn is predicted to be a double acceptor
with two hyperdeep levels, suggesting poor prospects for creating the
delocalized holes in the valence band that are required to mediate a
ferromagnetic interaction within the Zener model.  Explicit
calculation of the effective Heisenberg coupling between Mn in nearby
divacancy sites confirms that Mn-doped diamond shows little promise of
ferromagnetism above a few Kelvins.

Our findings may not be unique to diamond: other semiconductors with
predicted high Curie temperatures have similarly small atomic volumes,
which may likewise make the substitutional site unfavorable for Mn
doping.  Comprehensive calculations of the site energetics for GaN,
ZnO, and other host semiconductors would be of great interest.

\section{Acknowledgements}

This work was supported by the Office of Naval Research and the DARPA
Spins in Semiconductors program.  Computations were performed at the
DoD Major Shared Resource Centers at ASC and NAVO.


\end{document}